# *Quantization of exciton charging in organic semiconductor*

*Dongcheng Chen,*[a] Yanfei Lu,[a] Qiaobo Wang,[b] Shi-Jian Su,[a]*

*[a]State Key Laboratory of Luminescent Materials and Devices and Institute of Polymer Optoelectronic Materials and Devices, South China University of Technology, Guangzhou 510640, P. R. China*

*[b]Oriental Spectra Technology (Guangzhou) Co., Ltd, Guangzhou 510651, P. R. China*

** E-mail: mschendc@scut.edu.cn*

An exciton is a quasiparticle that are made of a bounded electron-hole pair, which can exist for certain lifetime. Formation of room-temperature excitons is a unique characteristic of amorphous bulky organic semiconductors in comparison to inorganic counterparts. The permittivity of organic semiconductors is generally low, because of the loose molecular packing. The low permittivity leads to large Coulombic force between the electron and hole in pair, resulting in existence of excitons at room temperature. One exciton can be charged by additional positive or negative charges, giving a many-body quasiparticle. This is exciton charging effect. Exciton charging was reported in two-dimensional transition metal dichalcogenides (TMDs) inorganic semiconductors[1-3]. The two-dimensional nature of TMDs is the intrinsic factor that leads to the anisotropic screening and strong in-plane Coulombic interaction.[4] This gives rise to the formation of exciton-relevant quasiparticles, like trions and biexcitons. It is reasonable to guess that exciton charging might also occur in organic semiconductors due to the strong exciton property. The bulky organic semiconductors consist of conjugated molecules by Van der Waals' force, which, in principle, allows being of exciton charging in three-dimensional morphology. The interaction between excitons and charges (or polarons) is widely believed to influence the properties of organic optoelectronic devices. For example, the exciton-polaron annihilation was found to lead to the efficiency-roll,[5-6] and aging of the devices.[7-8] Besides, the trions were observed in organic semiconductors under magnetic field, which is adopted as the particle model to interpret the magnetoresistance phenomena.[9-10] However, the quasiparticle, directly formed from exciton charging in three dimensional organic semiconductors has not been observed before.

In this work, exciton charging in bulky amorphous organic molecule semiconductor was observed and more interestingly, we found that exciton charging is quantized which is dependent on the current density through the semiconductor. The exciton charging leads to generation of new quasiparticles in semiconductor at room temperature. The term, excion (from the etyma of exciton and ion), was proposed to name these kinds of quasiparticles for convenience. The positive excions were found to drift through the semiconductor more coherently than free holes. This work might suggest the possibility of developing novel optoelectronic devices based on the quasiparticle from exciton charging.

The model organic semiconductor investigated here is N,N′-Di (1-naphthyl)-N,N′-diphenyl- (1,1′-biphenyl)-4,4′-diamine (NPB). NPB is a widely used hole transporter in the field of organic light-emitting diodes. The evaporated thin film of NPB is amorphous. Two devices with a layer structure of ITO/NPB(1208.3 nm)/Al(120 nm) were investigated, where the device without and with a $O_2$ plasma treatment on ITO surface is denoted as A and B, respectively. The $O_2$ plasma treatment can increase the work function of indium tin oxide (ITO) electrode thus can facilitate hole injection from ITO to NPB layer. The highest occupied molecular orbital (HOMO) level of NPB is -5.5 eV, and lowest unoccupied molecular orbital (LUMO) level is -2.4 eV. Work function of ITO without $O_2$ plasma treatment is around -4.6 eV, and after treatment the value is around -5.2 eV. Work function of Al is around -4.2 eV, which can build up a big electron injection barrier for NPB/Al interface.

A forward bias, i.e., ITO acts as the anode and Al serves as the cathode, of the device A and B at the same voltage, will lead to distinct hole volume current density. **Fig. 1** displays the volume current density versus the voltage of device A and B. The injected current density of device B is much higher than A. In the test range, the injection current of device A is nearly negligible. The $O_2$ plasma treatment leads to a smaller hole injection barrier for B, thus much higher hole current can be achieved. We will discuss later that it is the difference in steady background currents of device A and B that leads to completely different characteristics upon transient optical excitation.

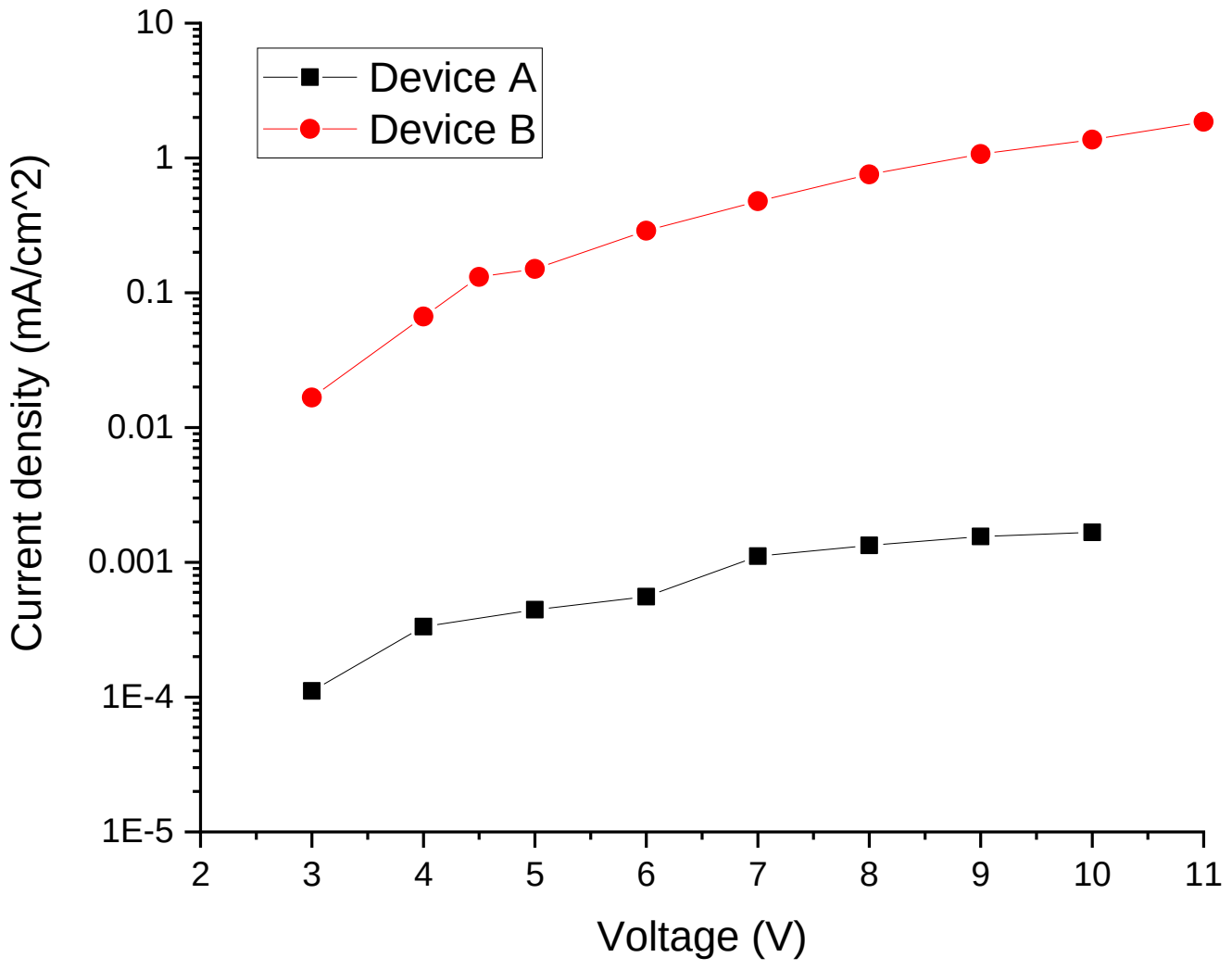

**Fig. 1** The logarithm of stationary current density versus the voltage of the device A and B without optical excitation.

We further investigated the properties of device A and B by a time-of-flight measurement system. A nitrogen transient laser with a wavelength of 337.1 nm was used as the excitation source. During the measurement, the pulse energy is constant. A lump of excitons was generated within the duality of laser beam. A different electric field was applied to each sample. The photocurrent was recorded from the sampling resistor by the oscilloscope. **Fig. 2** shows the transient photocurrent versus time of device A under various electric field. At an electric field below $3*10^6$ V·m$^{-1}$, the photocurrent is small. From $3.27*10^6$ to $8.36*10^6$ V·m$^{-1}$, the photocurrent is getting larger and larger. The falling edge of the photocurrent is significantly affected by the electric field. The bigger the electric field is, the smaller time the falling edge occurs. Distinct crossover points between the curves at higher and lower electric field can be observed, indicating that the falling edge results from the quenching of holes at the cathode. From the turning point of the falling edge, we extracted the transient time and calculated the mobility of NPB by the following equation: $\mu = \frac{d^2}{V \cdot t_T}$, $d$ is the thickness of the film, $V$ is the applied voltage across the device and sampling resistor, $t_T$ is the flight time. As shown on the inset of **Fig. 2**, the mobility varies from $6.83*10^{-8}$ to $9.14*10^{-8}$ $m^2 \cdot V^{-1} \cdot s^{-1}$) under an electric field in the range of $3.27*10^6$ to $8.36*10^6$ $V \cdot m^{-1}$. These mobility values conform well with reported results of NPB by the TOF method.[11-12] Since the coupling mode for the signal input to oscilloscope is direct coupling (i.e., without the chopass), the current before the time zero directly reveal the injected steady current across the device. For the measured electric field, the injected steady current is small and nearly the same, which is consistent with results in Fig. 1.

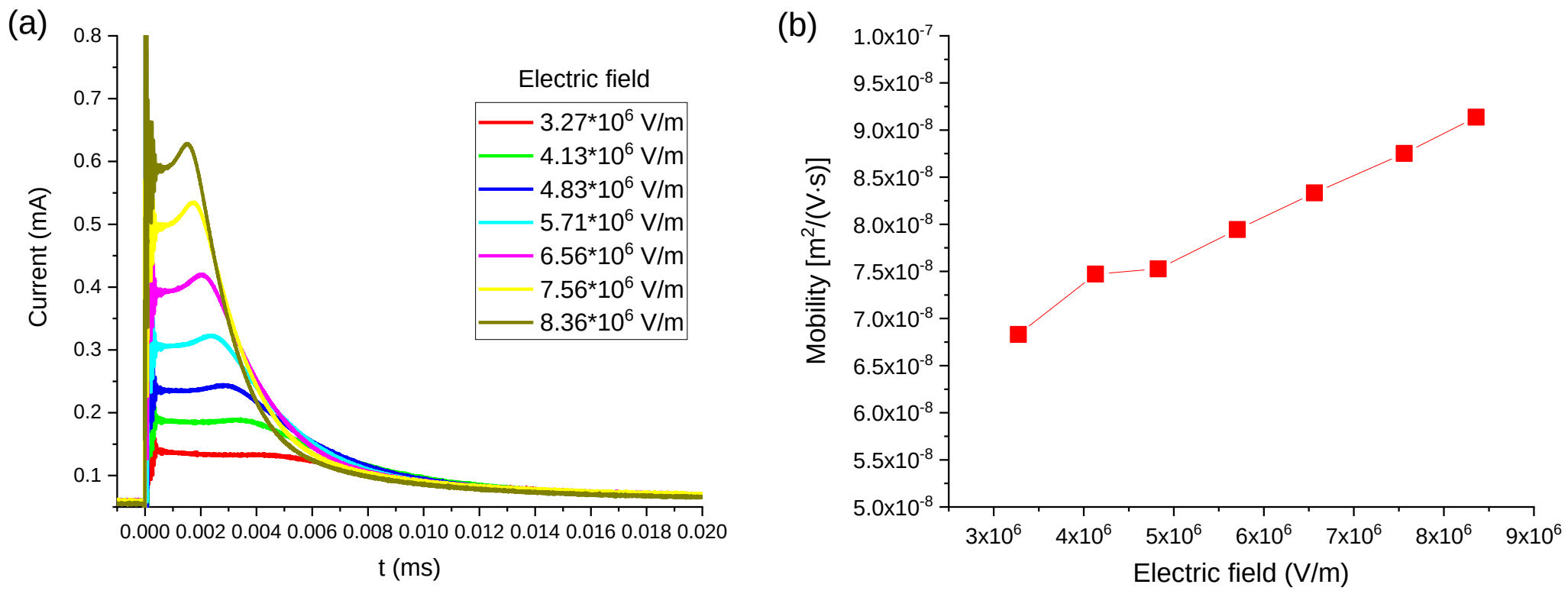

**Fig. 2** (a) photocurrent of device A in the function of different electric fields; (b) the hole mobilities of NPB from device A under different electric fields.

The response of device A can be well explained by the following analysis. Under the bias, the net electric force of the excitons is zero, due to the neutrality of excitons. Whereas the electric field generates two consequences on the excitons: 1) the torque on the exciton is introduced, which give rise to the rotation of excitons and orientation parallel to the electric field; 2) the exciton is stretched by the electric field and eventually dissociates into free charge. Under the forward bias, the electrons move and are quenched by the anode quickly, while the holes drift towards the cathode. Since the charges are generated close to the anode, the electron current disappears quickly and cannot be observed for our measurement system. When the drifted holes reach the cathode, then they are quenched, giving the turning point, i.e., the point to determine $t_T$, at the start of the falling edge. The shape of the time dependent transient photocurrent is affected by concentration of photo-induced charges, electric field, and traps density, which will be discussed by our later work.

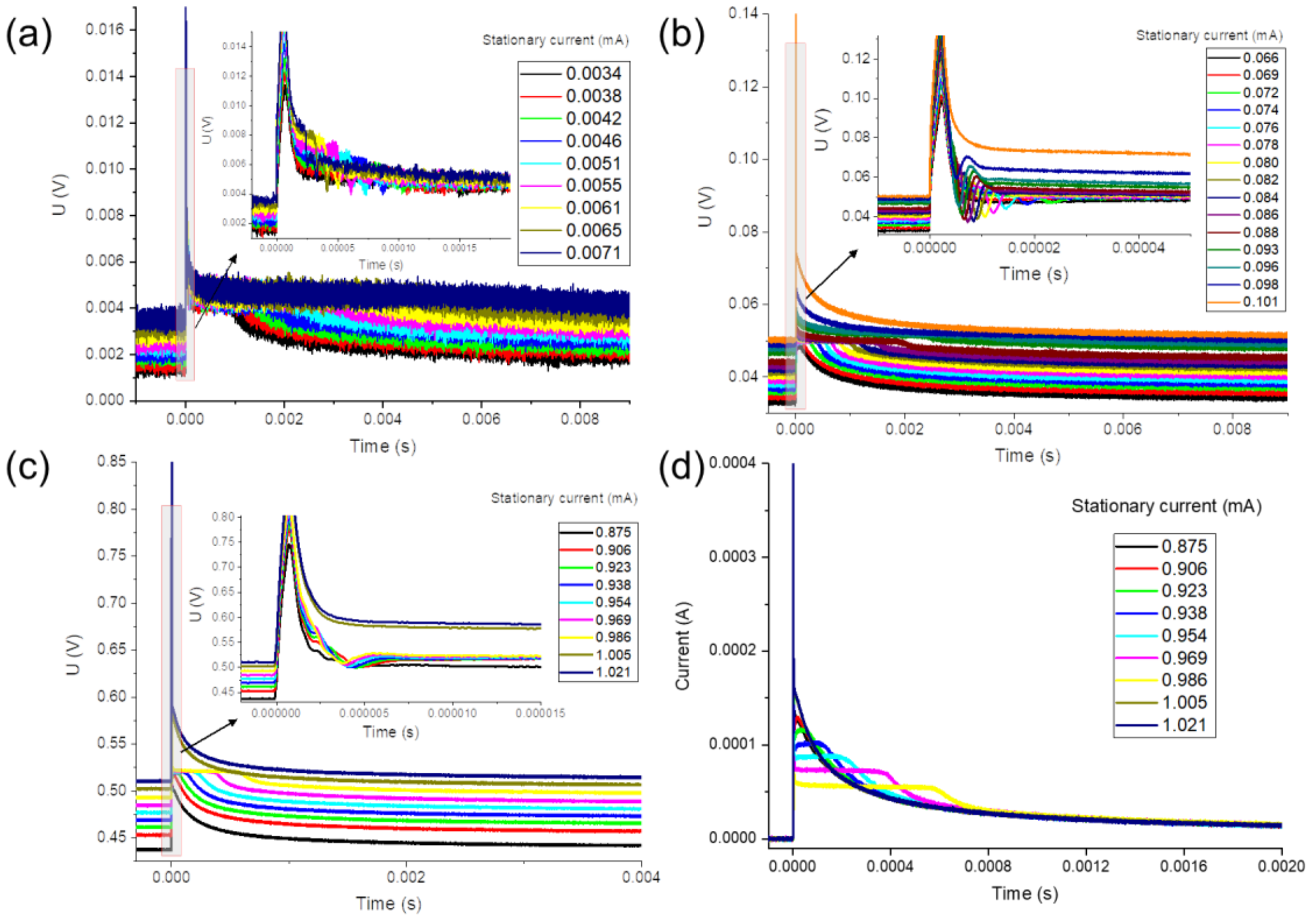

**Fig. 3** The temporal photocurrent of Device B under different electric fields which corresponds the stationary currents from 0.0034 to 0.0071 mA (a), 0.066 to 0.101 mA (b), and 0.875 to 1.021 mA (c). Insets display the enlarge details to clearly show the exciton charging points. (d) the transient photocurrent which is got by subtracting the common mode of the data in (c).

For device B, we observed distinct phenomenon under different electric fields with increased steady hole current in contrast to device A. The electric field applied to device B was changed in a step of 827.6 V·cm$^{-1}$ (corresponding to the voltage of 0.1 V) from zero to a value that leads to the device burnout. With the increase of electric fields, we could clearly identify different characteristics of the transient current. For easy visibility, the data in three ranges of the different stationary currents, i.e., corresponding to various electric fields, that gives unusual signals, are shown in **Fig. 3** (a), (b), (c), respectively. The other data not shown here mainly contains the information of so-called Zone I and Zone IV later, which can be understood similarly as Device A. Four time zones with different characteristics can be roughly classified for the data in **Fig. 3** (a), (b), (c). In zone I, the photocurrent characteristic is from the generation and decay of the photo-induced charges arising from the separation of excitons. And this origin is similar as Device A. In zone II, for some cases of certain stationary current, clear quenching points of transient current were observed. In zone III, the transient characteristic behaves like a non-dispersion plateau, which is similar as the non-dispersive case in conventional TOF measurement. In zone IV, it is the falling edge and the turning point at the start of the falling edge gives the charge quenching point by the counter electrode. The experiment with the same step change of the electric fields to device A was also performed, however, similar features were not seen. Since the main difference between device A and B is the steady electro-injected current density, we considered the occurrence of these unique characteristics are dependent on the stationary current that are across the device. To verify this, the resistance of sampling resistor was changed from 50 to 3000 Ω to test if the electric field or the current is the key factor. Since the internal resistance of the sample is very big, the electric field across the sample stays nearly identical with the changing sampling resistance. We observed that with the tuned sampling resistance, the non-dispersive feature was greatly different (**Fig. 4**). This leads us to conclude that it is the steady current that gives rise to the non-dispersive zone in device B. These features of device B are distinctly different from device A by the following aspects: 1) the range of electric fields that give full characteristics from Zone I to IV occurs periodically; 2) the length of Zone III increases with the electric fields in the non-dispersive zone; 3) each non-dispersive zone is quite narrow, which can be easily neglected if the step of electric field is big; 4) at the start edge of the plateau, a transient point that decreases in time with the increase of the electric field could be identified; 5) the profile of the photocurrent is very sensitive to the current density through the device.

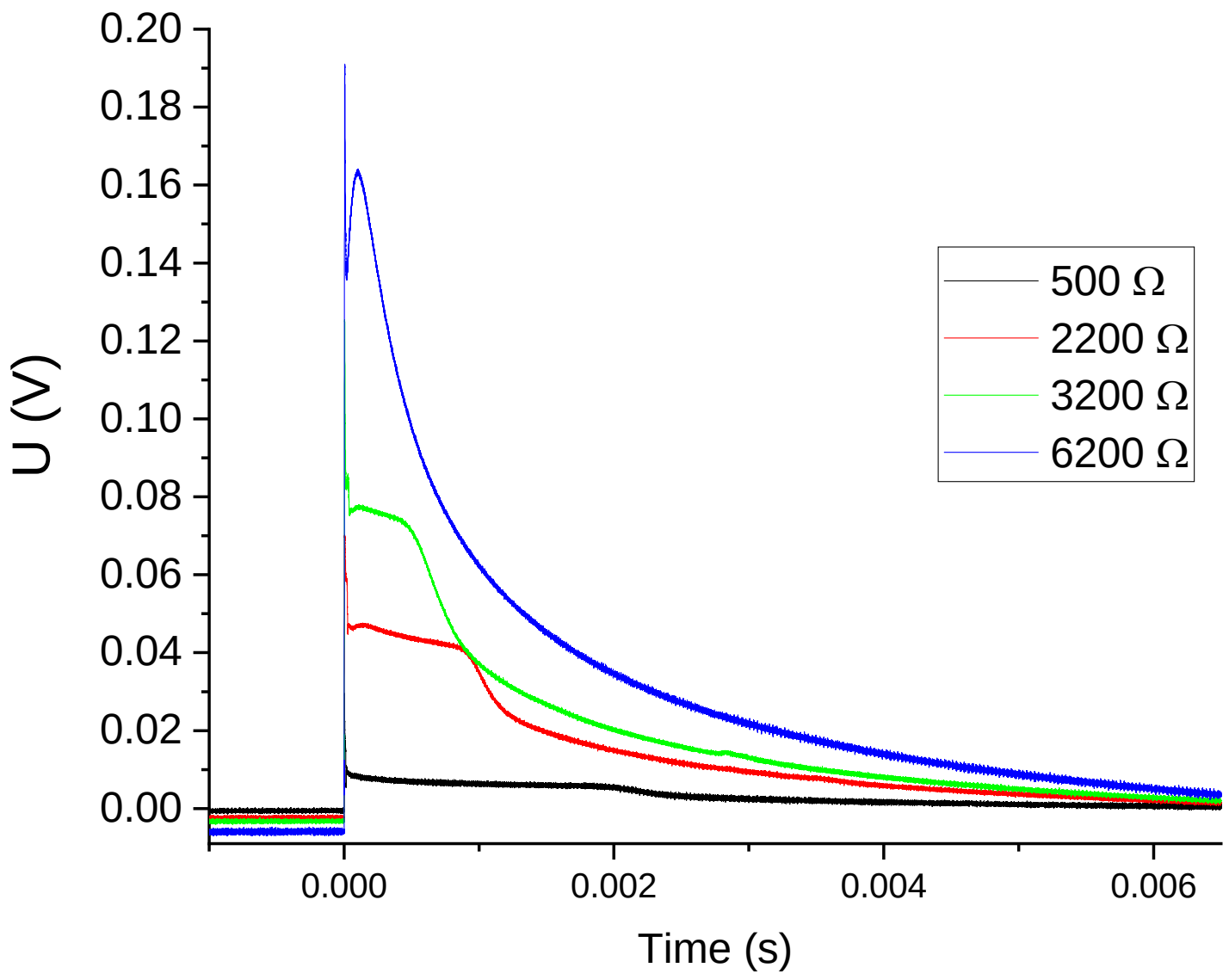

**Fig. 4** Photocurrent of Device B at an electric field of $8.19*10^4$ V $cm^{-1}$ probed with different sampling resistance.

It is obvious that the analysis of the electrons dynamics in device A is not suitable for interpreting what it is observed for device B. From the previous analysis of device A, it can be understood that the free holes are not directly responsible for the photocurrent characteristic of device B. Since it is the charge species that lead to generation of drift current, the exciton polariton, which is formed from the coupling of the exciton and electromagnetic mode and is of electric neutrality, can be ruled out. A resonant cavity with a good quantity factor is also absent in our device, which also doesn't support the involvement of exciton polaritons. We proposed that it is the exciton charging that leads to the characteristics of device B. A direct and strong evidence for the existence of exciton charging can be identified from the charging points of the transient photocurrent curves. As shown in the inset of Figure 3 (a), (b), (c), clear charging points can be observed for the exciton charging region. It can also be seen that the photocurrent curves outside the exciton charging region don't display the charging points. The larger the stationary current as well as the electric field is, the smaller charging point in time is. It is interesting that the exciton charging occurs in a way of quantization, which is dependent on the stationary current through the device. For device A, the injected stationary current density is nearly negligible, thus the exciton charging effect was not found.

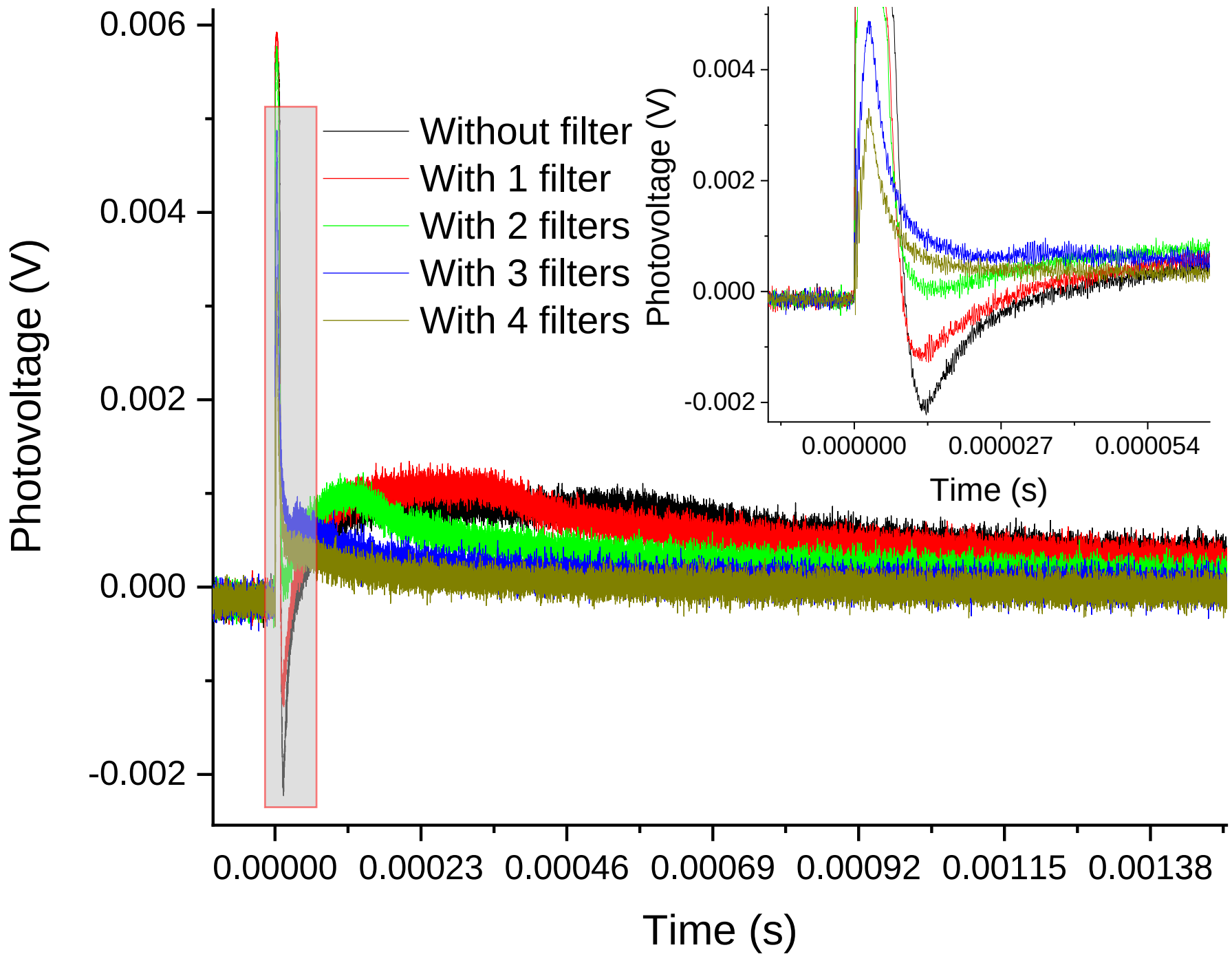

**Fig. 5** The transient photocurrent of device B probed at the AC coupling mode and at a constant stationary current, whereas, under the different intensities of pump light by adding various number of the same filters.

Another experiment to support the exciton charging explanation was done by changing the intensity of the excited light. In this experiment, the input coupling mode of the oscilloscope was changed from DC mode to AC mode to filter the low frequency stationary component, thus, to better manifest the temporal signal of exciton charging. The stationary current for this test remains at 98 μA, while the intensity of excited light was changed by adding various number of filters. The results are shown in **Fig. 5**. Without the filter, the device with 98 μA stationary current exhibits clear characteristics of exciton charging and drift of excions. Much more pronounced exciton charging point can be identified for this test, due to the exclusion of the common mode of stationary current. With lower intensity of the excited beam, it was observed that the drift feature decreases and finally the feature of exciton charging totally disappears for the measurement with 4 filters. The profile of the conditions with 4 filters is a conventional peak without any quenching and non-dispersive characteristics. This can be well explained by lacking enough exciton concentration due to a weak pump beam. This suggests that the exciton charging needs to be triggered by an enough exciton concentration. In overall, the above experiments strongly support that the occurrence of exciton charging is dependent on an appropriate concentration of excitons and a suitable magnitude of stationary current through the device.

In summary, we comparatively investigated two devices with the same device configuration, however, totally different injected stationary current due to different anode surface treatment. Distinctly different transient characteristics were observed for these two devises. It was proposed that it is the exciton charging that leads to the unique property of the latter device with easy electro-injected hole current. It was observed that the exciton charging occurs in a quantized manner, which is dependent on the magnitude of the background stationary current. We also proposed for convenience, the concept, excions, can be used for naming the quasiparticles that are generated from exciton charging. More experiments to further clarify the property of excions are urgently needed, and development of novel optoelectronic devices by excions in molecular organic semiconductors is possible and potential. Besides, the investigation on excions might also help to understand certain properties of current well-known optoelectronic devices, like organic light-emitting diodes and solar cells.